\documentclass[aps,prd,reprint,preprintnumbers,superscriptaddress,showpacs,twocolumn]{revtex4-1}
\usepackage{latexsym,graphicx,amssymb,amsmath,mathrsfs}
\usepackage{setspace,bm}
\usepackage[breaklinks, colorlinks=true, pdfstartview=FitV, linkcolor=red, citecolor=blue, urlcolor=blue]{hyperref}
\usepackage[usenames]{color}
\usepackage[normalem]{ulem}
\usepackage{latexsym}
\usepackage{epstopdf}
\usepackage{mathtools}
\usepackage{url}
\usepackage{comment}
\usepackage{braket}
\usepackage{CJKutf8}

\renewcommand\sout{\bgroup \color{red} \ULdepth=-.5ex \ULset}

\newcommand{\bequ}{\begin{equation}}
\newcommand{\eequ}{\end{equation}}
\newcommand{\bea}{\begin{eqnarray}}
\newcommand{\eea}{\end{eqnarray}}

\DeclareSymbolFont{boldletters}{OML}{cmm} {b}{it}
\DeclareSymbolFontAlphabet{\mathbit}{boldletters}
\DeclareMathSymbol{\alpha}{\mathalpha}{letters}{"0B}
\DeclareMathSymbol{\beta}{\mathalpha}{letters}{"0C}
\DeclareMathSymbol{\gamma}{\mathalpha}{letters}{"0D}
\DeclareMathSymbol{\delta}{\mathalpha}{letters}{"0E}
\DeclareMathSymbol{\epsilon}{\mathalpha}{letters}{"0F}
\DeclareMathSymbol{\zeta}{\mathalpha}{letters}{"10}
\DeclareMathSymbol{\eta}{\mathalpha}{letters}{"11}
\DeclareMathSymbol{\theta}{\mathalpha}{letters}{"12}
\DeclareMathSymbol{\iota}{\mathalpha}{letters}{"13}
\DeclareMathSymbol{\kappa}{\mathalpha}{letters}{"14}
\DeclareMathSymbol{\lambda}{\mathalpha}{letters}{"15}
\DeclareMathSymbol{\mu}{\mathalpha}{letters}{"16}
\DeclareMathSymbol{\nu}{\mathalpha}{letters}{"17}
\DeclareMathSymbol{\xi}{\mathalpha}{letters}{"18}
\DeclareMathSymbol{\pi}{\mathalpha}{letters}{"19}
\DeclareMathSymbol{\rho}{\mathalpha}{letters}{"1A}
\DeclareMathSymbol{\sigma}{\mathalpha}{letters}{"1B}
\DeclareMathSymbol{\tau}{\mathalpha}{letters}{"1C}
\DeclareMathSymbol{\upsilon}{\mathalpha}{letters}{"1D}
\DeclareMathSymbol{\phi}{\mathalpha}{letters}{"1E}
\DeclareMathSymbol{\chi}{\mathalpha}{letters}{"1F}
\DeclareMathSymbol{\psi}{\mathalpha}{letters}{"20}
\DeclareMathSymbol{\omega}{\mathalpha}{letters}{"21}
\DeclareMathSymbol{\varepsilon}{\mathalpha}{letters}{"22}
\DeclareMathSymbol{\vartheta}{\mathalpha}{letters}{"23}
\DeclareMathSymbol{\varpi}{\mathalpha}{letters}{"24}
\DeclareMathSymbol{\varrho}{\mathalpha}{letters}{"25}
\DeclareMathSymbol{\varsigma}{\mathalpha}{letters}{"26}
\DeclareMathSymbol{\varphi}{\mathalpha}{letters}{"27}
\DeclareMathSymbol{\Gamma}{\mathalpha}{letters}{"00}
\DeclareMathSymbol{\Delta}{\mathalpha}{letters}{"01}
\DeclareMathSymbol{\Theta}{\mathalpha}{letters}{"02}
\DeclareMathSymbol{\Lambda}{\mathalpha}{letters}{"03}
\DeclareMathSymbol{\Xi}{\mathalpha}{letters}{"04}
\DeclareMathSymbol{\Pi}{\mathalpha}{letters}{"05}
\DeclareMathSymbol{\Sigma}{\mathalpha}{letters}{"06}
\DeclareMathSymbol{\Upsilon}{\mathalpha}{letters}{"07}
\DeclareMathSymbol{\Phi}{\mathalpha}{letters}{"08}
\DeclareMathSymbol{\Psi}{\mathalpha}{letters}{"09}
\DeclareMathSymbol{\Omega}{\mathalpha}{letters}{"0A}

\begin{document}
\title{Hadron-quark transition and chiral symmetry restoration at high density} 

\author{Hiroaki Kouno}
\email[]{kounoh@cc.saga-u.ac.jp}
\affiliation{Department of Physics, Saga University,
             Saga 840-8502, Japan}

\author{Kouji Kashiwa}
\email[]{kashiwa@fit.ac.jp}
\affiliation{Fukuoka Institute of Technology, Wajiro, Fukuoka 811-0295, Japan}


\begin{abstract}
A simple phenomenological hybrid hadron-quark model with effective volume effects of baryons and chiral dynamics is investigated.  
The hybrid EoS naturally connects the low-density baryonic matter with the high-density quark matter. 
In the intermediate region, a phase, which cannot be regarded as pure hadron matter or pure quark matter, appears.
In this model, there is a possibility that the abrupt first-order-like transition to pure quark matter induces the strong chiral symmetry restoration and the speed of sound has a large peak at considerable large density. 
\end{abstract}

\maketitle


\section{Introduction}

Exploration of the QCD phase diagram is an important subject not only in particle and nuclear physics but also in astrophysics and cosmology; for a review, see, e.g., Ref.~\cite{Fukushima:2010bq} and references therein.  
However, at finite baryon (or quark) chemical potential, the first principle calculation, i.e., the lattice QCD simulation, is not feasible due to the infamous sign problem. 
To avoid the problem, several methods are proposed and investigated, but, at present,  these methods are not complete, and we do not have the established equation of state (EoS) at finite baryon density. 
 
At low temperature, there is nuclear (baryonic) matter at saturation density. 
As the density increases, other baryons may appear.  
At the extreme high density, the chiral symmetric quark matter is expected to appear and conformality is restored.  
However, at present, we do not have a definite information of the
EoS in the intermediate region between the saturation density and the extremely high density. 

It is known that repulsive effects among baryons are important in the intermediate region. 
If repulsion is absent, the baryonic matter is realized at sufficiently large baryon density~\cite{Cleymans:1985wb}. 
One of the traditional treatments of such repulsion is to consider the excluded volume effects (EVE) among baryons~\cite{Cleymans:1986cq,Kouno:1988bi,Rischke:1991ke}.  
The excluded volume effects successfully prevent baryonic matter from realizing at sufficiently large baryon density; for a recent review, see, e.g., Ref.~\cite{Fujimoto:2021dvn} and references therein. 
 
At large density, the chiral symmetry restoration is also expected. 
The Nambu--Jona-Lasinio (NJL) model~\cite{Nambu:1961tp,*Nambu:1961fr} is a simple but very useful model to describe the restoration; as a review, see, e.g., Ref.~\cite{Hatsuda:1994pi} and references therein. 
However, the NJL model cannot describe the hadron-quark transition.  

Furthermore, recently, it has been emphasized that the trace anomaly and the speed of sound are very important to understand the properties of the high density hadron and quark matter; 
see, e.g., Refs.~\cite{Fujimoto:2022ohj,Kojo:2020krb} and references therein. 

In this paper, we construct a simple phenomenological hybrid hadron-quark model with EVE of baryons and chiral dynamics.  
The model naturally connects the baryonic matter at low density and the quark matter at high density. 
It can also describe the chiral restoration. 
It is found that, in this hybrid model, there is a possibility that the abrupt first-order like transition to pure quark matter induces the strong restoration of chiral symmetry and the speed of sound has a large peak.

This paper is organized as follows. 
In Sec.~\ref{formalism}, the hybrid model is formulated. 
In Sec.~\ref{Nresults}, numerical results are shown for two typical cases. 
Section~\ref{summary} is devoted to a summary and discussion.

\section{Formalism}
\label{formalism}

First, we give a sketch of our strategy of calculations.  
Our main purpose is to know the $\mu_{\rm B}$-dependence of the thermodynamic quantities of hadron or quark matter, where $\mu_{\rm B}$ is the baryon chemical potential. 
First, we construct the baryon number density $n_{\rm B}$ with the EVE of baryons. Next we require that $n_{\rm B}$ approaches ${n_{\rm q}\over{3}}$ in the high density limit, where $n_{\rm q}$ is the quark number density of pure quark phase. In this procedure, $n_{\rm B}$ depends on the chiral condensates $\sigma_f~(f=u,d,s)$ which are included in the quark model. 
Using the thermodynamic equation, the $\sigma_f$-dependence of the thermodynamic potential density $\Omega~(=-P)$ is obtained, where $P$ is the pressure of the system.  
The values of $\sigma_f$ are determined to minimize $\Omega$ (or maximize $P$). Using the obtained values of $\sigma_f$, the other thermodynamic quantities are calculated. 

\begin{figure}[h]
\centering
\centerline{~~~~~~~~\includegraphics[width=0.35\textwidth]{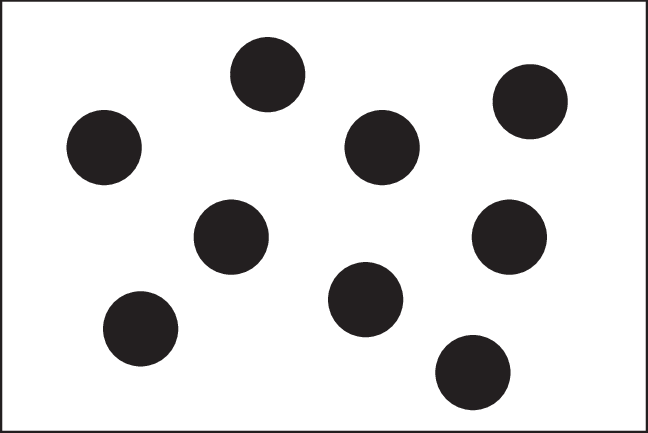}}
\bigskip
\caption{Schematic figure of $N$ non-pointlike baryons 
in the system with volume $V$. 
 }
 \label{Fig_volume1}
\end{figure}

\begin{figure}[h]
\centering
\centerline{~~~~~~~~\includegraphics[width=0.35\textwidth]{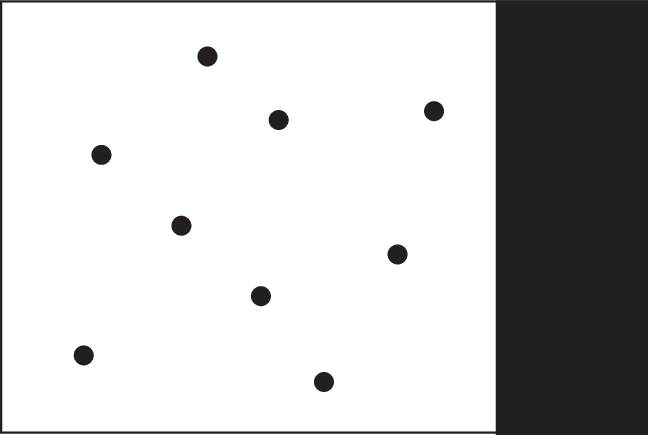}}
\bigskip
\caption{Schematic figure of $N$ pointlike baryons 
in the system with effective volume $V-v_{\rm B}N$. 
The volume of the dark 
region is $v_{\rm B}N$.  
 }
 \label{Fig_volume2}
\end{figure}

As shown in Fig.~\ref{Fig_volume1}, we consider $N$ non-pointlike baryons in the system with volume $V$ where $N$ is the number of baryons.   
We consider this system to be equivalent to the system of $N$ pointlike baryons in the effective volume $V-v_{\rm B}N$ where $v_{\rm B}$ is the volume of a baryon; see Fig.~\ref{Fig_volume2}.   
Then, the following equation is satisfied; 
\begin{eqnarray}
\tilde{n}_{\rm B}={N\over{V-v_{\rm B}N}}={n_{\rm B}\over{1-v_{\rm B}n_{\rm B}}},  
\label{number_ex_f1}
\end{eqnarray}
where $n_{\rm B}={N\over{V}}$ is the baryon density of non-pointlike baryon and $\tilde{n}_{\rm B}$ is the one of $N$ pointlike baryons. 
In this paper, we add a tilde to the baryon number density of point-like baryons. 
Hence, the baryon number density of baryonic matter with EVE is given by 
\begin{eqnarray}
n_{\rm B}={\tilde{n}_{\rm B}\over{1+v_{\mathrm B}\tilde{n}_{\rm B}}}< {1\over{v_{\rm B}}}.  
\label{number_ex}
\end{eqnarray}

\begin{figure}[h]
\centering
\centerline{~~~~~~~~\includegraphics[width=0.35\textwidth]{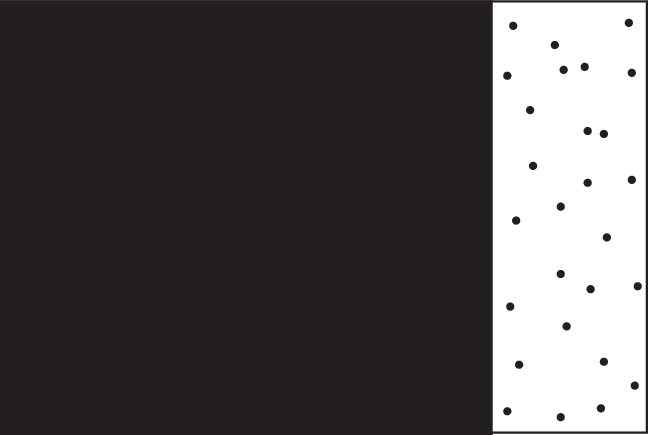}}
\bigskip
\caption{Schematic figure of $3N$ quarks in the system with effective volume $v_{\rm B}N$. 
The volume of the 
dark region is $3v_{\rm q}N$.  
 }
 \label{Fig_volume3}
\end{figure}

The pressure $P$ of the baryonic matter with the excluded volume effects is determined by the thermodynamic equation
\begin{eqnarray}
{\partial P (T,\mu_{\rm B})\over{\partial \mu_{\rm B}}}=n_{\rm B}.
\label{pressure_ex}
\end{eqnarray}
The other thermodynamic quantity, such as the energy density $\varepsilon$, is determined by the thermodynamic relation.  
Hereafter, we concentrate on EoS at zero temperature and omit the variable $T$ for simplicity. 
Then, Eq.~(\ref{pressure_ex}) is represented as ${dP\over{d\mu_{\rm B}}}$. 

When a constant $v_{\rm B}$ is used, the speed of sound, 
\begin{eqnarray}
c_{\rm Bs}=\sqrt{{d P_{\rm B}\over{d \varepsilon_{\rm B}}}}=\sqrt{\left.{d P_{\rm B}\over{d\mu_{\rm B}}}\right/{d \varepsilon_{\rm B}\over{d\mu_{\rm B}}}},
\label{speed_s}
\end{eqnarray}
may exceed the speed of light $c~(=1)$ and the causality can be easily violated.  
In fact, when $v_{\rm B}\tilde{n}_{\rm B}\gg 1$, we have 
\begin{align}
&\frac{dP_{\rm B}}{{d\mu_{\rm B}}}
= n_{\rm B}
= \frac{\tilde{n}_{\rm B}}{{1+v_B\tilde{n}_{\rm B}}}
\sim \frac{1}{{v_{\rm B}}},
\label{number_ex_2a}\\
&{d\varepsilon_{\rm B}\over{d\mu_{\rm B}}}={d\over{d\mu_{\rm B}}}\left(\mu_{\rm B}n_{\rm B}-P_{\rm B}\right)
=\mu_{\rm B}{dn_{\rm B}\over{d\mu_{\rm B}}}\sim 0. 
\label{number_ex_2}
\end{align}
This means that the EoS becomes very hard and the speed of sound diverges.
Therefore, the density dependence of $v_{\rm B}$ is very important. 

It is natural to assume that the baryon number of the system approaches the one of the pure quark system with the same chemical potential, when the baryon chemical potential is very large. 
This requirement can be achieved if we assume 
\begin{eqnarray}
v_{\rm B}={3\over{n_{\rm q}}}~~~~
(n_{\rm q}\neq 0), 
\label{volume_1}
\end{eqnarray}
where $n_{\rm q}$ is the quark number density of the pure quark matter. 
In other words, we assume that EoS inside the baryons is described by the pure-quark model. 
The baryon number density is given by 
\begin{align}
n_{\mathrm B}
&= \frac{\tilde{n}_{\mathrm{B}}}
       {{1 + v_{\mathrm B}\tilde{n}_{\mathrm{B}}}}
\label{number_ex_3_c}
\\
&= \frac{n_{\mathrm q}\tilde{n}_{\rm B}}
        {{n_{\mathrm q}+3\tilde{n}_{\rm B}}}
< \frac{n_{\mathrm q}}{{3}}.
\label{number_ex_3}
\end{align}
Hence, the EoS is expected to approach the pure quark EoS at high density. 
When $3\tilde{n}_{\rm B}\gg n_{\rm q}$, we obtain 
\begin{eqnarray}
n_{\rm B}\sim {n_{\rm q}\over{3}}.    
\label{number_ex_4}
\end{eqnarray}
When $n_{\rm q}\neq 0$ and $\tilde{n}_{\rm B}\neq 0$, Eq. (\ref{number_ex_3}) can be rewritten as 
\begin{eqnarray}
n_{\rm B}&=&{{1\over{3}}n_{\rm q}\over{1+v_{\rm q}n_{\rm q}}}
\label{number_ex_3_b}
\\
&=&{1\over{v_{\rm B}+3v_{\rm q}}}
\label{number_ex_3_d}
\\
&=&{1\over{3}}\left({1\over{n_{\rm q}}}+{1\over{3\tilde{n}_{\rm B}}}\right)^{-1},
\label{number_ex_3_e}
\end{eqnarray}
where $v_{\rm q}={1\over{3\tilde{n}_{\rm B}}}$. 
From Eq.~(\ref{number_ex_3_b}), we see that the system can also be regarded as the matter of quarks with finite effective volume $v_{\rm q}$; see Fig.~\ref{Fig_volume3}.  
In this sense, this model has a quark-hadron duality; for a general review of quark-hadron duality, see, e.g., Ref.~\cite{Shifman:2000jv} and references therein. 
It is also interesting that Eq.~(\ref{number_ex_3_e}) resembles the law of combined resistance of parallel resistances. 
The large $v_{\rm B}$ causes the strong suppression of baryonic matter. 
When $3v_{\rm q} < v_{\rm B}$, it is natural to regard that the system is composed of quarks with small volume $v_{\rm q}$ rather than baryons with large volume $v_{\rm B}$.

However, when $n_{\rm q}=0$, $v_{\rm B}$ cannot be defined.  
On the other side, it is known that, below the saturation density, the nucleon has a finite volume $v_{\rm B0}={4\pi r_{\rm B0}^3\over{3}}$ with the nucleon radius $r_{\rm B0}= 0.8$~fm. 
Hence, we interpolate between $v_{\rm B0}$ and ${3\over{n_{\rm q}}}$ with the following smooth function of $n_{\rm q}$; 
\begin{align}
v_{\rm B} &= {3\over{n_{\rm q}^\prime }},
\label{volume_3}
\end{align}
with
\begin{align}
n_{\rm q}^\prime
&= n_{\rm q}
 + \frac{3}{{v_{\rm B0}}}
   \exp{ \Bigl[ -a\Bigl( \frac{v_{\rm B0} n_{\rm q}}{3} \Bigr)^2 \Bigr]},
\label{pq_prime}
\end{align}
where $a$ is a free parameter and controls how abruptly $v_{\rm B}$ approaches $3/n_{\rm q}$. 
Note that, as is shown below, $n_{\rm q}$ depends on chiral condensates in our model. Hence, $v_{\rm B}$ and $n_{\rm B}$ also depend on the chiral condensates. 

For the pointlike hadron model, we use the hadron resonance gas (HRG) model. 
The number density is given as 
\begin{eqnarray}
\tilde{n}_{\rm B}=\sum_{i={\rm Baryon}} n_{\rm IFG}(g_{{\rm B}i}, m_{{\rm B}i}, \mu_{\rm B}), 
\label{n_HRG}
\end{eqnarray}
where $g_{{\rm B}i}$ and $m_{{\rm B}i}$ are the spin degeneracy and the mass of $i$-th baryon, respectively. 
The function $n(g, m, \mu)$ is the number density of ideal fermion gas with the degeneracy factor $g$, the mass $m$ and the chemical potential $\mu$ at zero temperature, and is given by 
\begin{eqnarray}
n_{\rm IFG}(g,m,\mu )=
\left\{
\begin{array}{cc}
0      &     (\mu <m) \\
\displaystyle
{g\over{6\pi^2}}(\mu^2-m^2) &  (\mu \geq m)
\end{array}
\right. .
\label{free_n}
\end{eqnarray}
In this paper, for simplicity, we use the same $v_{\rm B}$ for all baryons.  

For pure quark matter, we use the three-flavor NJL model with mean field approximation. 
The quark number density of the NJL model is given by 
\begin{eqnarray}
n_{\rm q}=\sum_{f=u,d,s} n_{\rm IFG}(6, M_f, \mu_{\rm q} ), 
\label{n_NJL}
\end{eqnarray}
where $M_f$ and $\mu_{\rm q} (=\mu_{\rm B}/3)$ are the effective mass of $f$-quark and the quark chemical potential, respectively. 
The effective quark mass is given by
\begin{eqnarray}
M_{f}&=&m_{f}-4G_{\rm s}\sigma_f+2G_{\rm d}\sigma_f^\prime\sigma_f^{\prime\prime}, 
\label{quakrmass}
\end{eqnarray}
with $f\neq f^\prime$, $f\neq f^{\prime\prime}$ and $f^\prime \neq f^{\prime\prime}$,  where $m_{f}$ is the current quark mass of $f$-quark, $G_{\rm s}$ and $G_{\rm d}$ are coupling constants of four and six-quarks interaction, and $\sigma_f$ is the chiral condensate of $f$-quark, respectively. 
In the pure quark system, the pressure is given by
\begin{eqnarray}
P=P_{\rm v}+P_{\rm D,NJL}-U_{\rm m},
\label{pressure_NJL}
\end{eqnarray}
where $P_{\rm v}$, $U_{\rm m}$ and $P_{\rm D,NJL}$ are the Dirac sea contributions, the mesonic and the density parts of the NJL model, respectively. 
For each flavor, the density part $P_{\rm D,NJL}$ is given by $P_{\rm F}(M_f,\mu_{\rm   q})$ which is equivalent to the pressure of the free quark gas with the quark mass $M_f$ and the quark chemical potential $\mu_{\rm q}$. 
The Dirac sea contributions are given by 
\begin{eqnarray}
P_{\rm v}(M_{f}) &=&-\sum_{f=u,d,s}{M_{f}^4\over{4\pi^2}}\left[
\left( {E_{{\rm v}f}\Lambda\over{M_{f}^2}}\right)\left( {E_{f}^2\over{M_{f}^2}}-{5\over{2}}\right)\right.
\nonumber\\
&&\left .+{3\over{2}}\log{\left( {E_{{\rm v}f}+\Lambda\over{M_{f}}}\right)}\right], 
\label{P_vacuum}
\end{eqnarray}
where $E_{{\rm v}f}=\sqrt{\Lambda^2+M_{f}^2}$ with the three dimensional momentum cutoff $\Lambda$. 
The Dirac sea contributions have no explicit $\mu_{\rm q}$-dependence. 
However, $P_{\rm v}$ depends on $\mu_{\rm q}$, since $M_{f}$ depends on $\mu_{\rm q}$. 
The mesonic part is given by
\begin{eqnarray}
U_{\rm m}=2G_{\rm s}(\sigma_u^2+\sigma_d^2+\sigma_s^2)-4G_{\rm d}\sigma_u\sigma_d\sigma_s. 
\label{U_mesonic}
\end{eqnarray}
According to Ref.~\cite{Rehberg:1995kh}, we set $m_{u,d}=5.5$ MeV, $m_s=140.7$ MeV, $G_{\rm s}\Lambda^2=1.835$, $G_{\rm d} \lambda^5=12.36$ and $\Lambda =602.3$ MeV. 

Instead of Eq.~(\ref{pressure_NJL}), in the hybrid model, the total pressure $P$ of the system is given by 
\begin{eqnarray}
P=P_{\rm v}+P_{\rm D}-U_{\rm m}.
\label{pressure_total}
\end{eqnarray}
Using $n_{\rm B}$ given by Eq.~(\ref{number_ex}) with Eq.~(\ref{volume_3}), the density part $P_{\rm D}$ is given by 
\begin{eqnarray}
P_{\rm D}(\mu_{\rm B},\sigma_f)=\int_0^{\mu_{\rm B}}d\mu \, n_{\rm B}(\mu,\sigma_f), 
\label{P_density}
\end{eqnarray}
where the integration is performed with fixing all $\sigma_f$ so as to satisfy the relation 
\begin{eqnarray}
\left.{\partial P(\mu_{\rm B},\sigma_f)\over{\partial \mu_{\rm B}}}\right|_{\sigma_f \mathrm{fixed}}=n_{\rm B}(\mu_{\rm B},\sigma_f).
\label{dP_n}
\end{eqnarray}
The value of the chiral condensate $\sigma_f$ is determined to maximize $P$ and satisfy 
\begin{eqnarray}
{\partial P(\mu_{\rm B},\sigma_f)\over{\partial \sigma_f}}=0.
\label{EOM}
\end{eqnarray}
Hence, the thermodynamic relation 
\begin{eqnarray}
{dP(\mu_{\rm B},\sigma_f(\mu_{\rm B}))\over{d\mu_{\rm B}}}&=&
\left.{\partial P\over{\partial \mu_{\rm B}}}\right|_{\sigma_f \mathrm{fixed}}+\sum_{f=u,d,s}{\partial P\over{\partial \sigma_f}}
{d\sigma_f\over{\mu_{\rm B}}}
\nonumber\\
&=&
\left.{\partial P_{\rm D}\over{\partial \mu_{\rm B}}}\right|_{\sigma_f \mathrm{fixed}}
\nonumber\\
&=&n_{\rm B}(\mu_{\rm B}, \sigma_f (\mu_{\rm B})),
\label{dP_n_2}
\end{eqnarray}
is satisfied. 
Therefore, using Eq. (\ref{P_density}), once  $P_{\rm D}$ is obtained as the function of $\sigma_f$ and the solution $\sigma_f$ which maximizes the total pressure (\ref{pressure_total}) is determined, the equation of motion (\ref{EOM}) and the thermodynamic relation (\ref{dP_n_2}) are automatically satisfied. 
The energy density $\varepsilon$ of the total system is determined using the thermodynamic relation
\begin{eqnarray}
\varepsilon =\mu_{\rm B}n_{\rm B}-P. 
\label{energy_density}
\end{eqnarray}
Following the above procedure, the quark and hadron matters are correlated with each other in the level of the pressure via EVE.
This is also true for the thermodynamic potential density $\Omega$ because there is a direct relation to the pressure, namely $\Omega =-P$.  
The functional form of $\Omega (\sigma_f)$ deviates largely from the one $\Omega_{\rm NJL}(\sigma_f)$ in the original NJL model in the intermediate region of $\mu_{\rm B}$. 
Therefore, our result can provide the crossover behavior even if the NJL model itself has the first-order chiral transition at low $T$ in the moderately high density region; since the thermodynamic potential is deformed by the hadron contributions, they act as the external term that explicitly breaks chiral symmetry. 

The hybrid model approaches in the NJL model in the high density limit. 
As is seen in the next section, when $n_{\rm B}$ approaches ${n_{\rm q}\over{3}}$ gradually, crossover chiral transition takes place. 
When $n_{\rm B}$ approaches ${n_{\rm q}\over{3}}$ abruptly, first-order like transition happens. 

The original NJL model has a cutoff. 
However, at zero temperature, the chemical potential is the natural cutoff for energy and momentum. 
Hence, in the numerical calculations, we use the cutoff only in the Dirac Sea contributions. 
In this procedure, the physical quantities are expected to approach those in the free quark gas model.  
In the HRG part, we include all baryons listed in the list of Particle Data Group~\cite{Workman:2022ynf}, but they only contribute to the results when their masses are smaller than the baryon chemical potential $\mu_{\rm B}$. 

Several hybrid models have been already proposed. 
For example, in Ref.~\cite{Masuda:2012ed}, the hyperbolic functions are used as an interpolation function of EoS. 
In our model, we do not give such an interpolation function but give a more microscopic quantity, namely, the density dependence of baryon volume in EVE. 
In Ref.~\cite{Jeong:2019lhv}, the hybrid model based on quarkyonic matter and the EVE is investigated. 
In our model, we do not assume quarkyonic matter, but some kind of more macroscopic quark-hadron duality (\ref{number_ex_3_b}) is assumed. 
The macroscopic model may be simpler and more convenient than the microscopic one, but it has a less dynamical description. The study of the relation between several hybrid models is an important problem in the future.    

\bigskip 

\section{Numerical results}
\label{Nresults}

Our hybrid model has one free parameter $a$ in Eq.~(\ref{pq_prime}). 
In this section, we show the numerical results in two typical cases, a crossover transition ($a=0.1$) and a first-order like transition ($a=0.8$).   
Hereafter, we use the quark chemical potential $\mu_{\rm q} ={\mu_{\rm B}\over{3}}$ instead of $\mu_{\rm B}$ since we are interested in the chiral dynamics of quarks as well as the hadron-quark transition. 

The numerical calculations are done according to the following procedures. 

~

\noindent
(1) For given $\mu_{\rm B}$, $\tilde{n}_{\rm B}$ is calculated using the HRG model. 

~

\noindent 
(2) For fixed value of $\sigma_f$, $n_{\rm q}$ is calculated. 

~

\noindent
(3) Using $\tilde{n}_{\rm B}$ and $n_{\rm q}$ as inputs, 
$n_{\rm B}$ is calculated as an output. 

~

\noindent
(4) Using Eq.~(\ref{P_density}) and ${n}_{\rm B}$, $\Omega$~($=-P$) is calculated for the fixed value of $\sigma_f$.

~

\noindent
(5) Changing the value of $\sigma_f$ and doing the same calculation as the procedure (2)$\sim$(4), a functional form $\Omega (\sigma_f)$ ($=-P(\sigma_f)$) is determined.  

~

\noindent
(6) Determine the solution $\sigma_f$ which minimizes $\Omega$ (maximizes $P$). 

~

\noindent
(7) Using the obtained solution $\sigma_f$, the other quantities are calculated as well as $P$ and $n_{\rm B}$. 

~

Note that the equation of motion (\ref{EOM}) and the thermodynamic relation (\ref{dP_n_2}) are automatically satisfied by minimizing (maximizing) $\Omega$ ($P$).

\subsection{Crossover transition}

\begin{figure}[t]
\centering
\centerline{\includegraphics[width=0.40\textwidth]{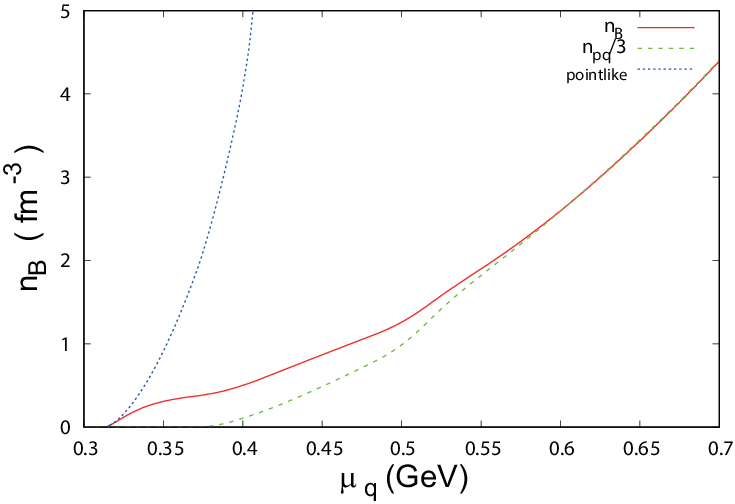}}
\caption{The $\mu_{\rm q}$-dependence of the baryon number density $n_{\rm B}$ with the crossover setup. 
The solid, dashed, and dotted lines show $n_{\rm B}$, ${n_{\rm q}\over{3}}$ and $\tilde{n}_{\rm B}$ (pointlike), respectively. 
 }
 \label{Fig_nB_010}
\end{figure}

First, we show the results with $a=0.1$; in the following, we call it the crossover setup. 
Figure~\ref{Fig_nB_010} shows the $\mu_{\rm q}$-dependence of the baryon number density $n_{\rm B}$. 
Note that not only $n_{\rm B}$ but also $n_{\rm q}$ depends on the values of the chiral condensates determined in the framework of the hybrid model. 
We see that $n_{\rm B}$ coincides with $\tilde{n}_{\rm B}$ at low density and, at $\mu_{\rm q}\sim 0.33$ GeV, begins to deviate from $\tilde{n}_{\rm B}$. 
As $\mu_{\rm q}$ increases, $n_{\rm B}$ approaches ${n_{\rm q}\over{3}}$ smoothly. 
Although the hadron-quark transition is a smooth crossover, we can regard the system as in the quark phase when $\mu_{\rm q} >0.55$ GeV. 
In the intermediate region $\mu_{\rm q}=0.33\sim 0.55$ GeV, it is difficult to identify the system as pure hadron matter or pure quark matter. 

\begin{figure}[t]
\centering
\centerline{\includegraphics[width=0.40\textwidth]{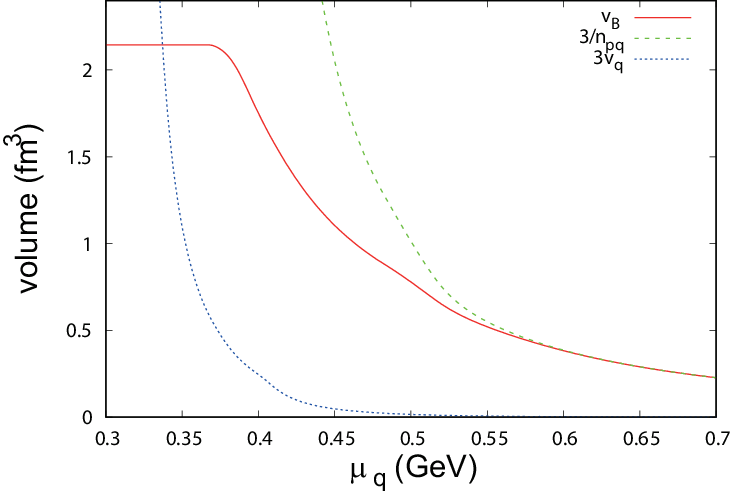}}
\caption{The $\mu_{\rm q}$-dependence of the baryon volume $v_{\rm B}$ with the crossover setup.  
The solid, dashed, and dotted lines show $v_{\rm B}$, ${3\over{n_{\rm q}}}$ and $3v_{\rm q}$, respectively. 
 }
 \label{Fig_volume_010}
\end{figure}

Figure~\ref{Fig_volume_010} shows the $\mu_{\rm q}$-dependence of the baryon volume $v_{\rm B}$. 
We see that $v_{\rm B}$ approaches $3/n_{\rm q}$ smoothly as $\mu_{\rm q}$ increases. 
When $\mu_{\rm q}>0.35$ GeV, $3v_{\rm q}$ is smaller than $v_{\rm B}$. 
Hence we may regard the system to be composed of quarks rather than baryons when $\mu_{\rm q} >0.35$ GeV. 

\begin{figure}[t]
\centering
\centerline{\includegraphics[width=0.40\textwidth]{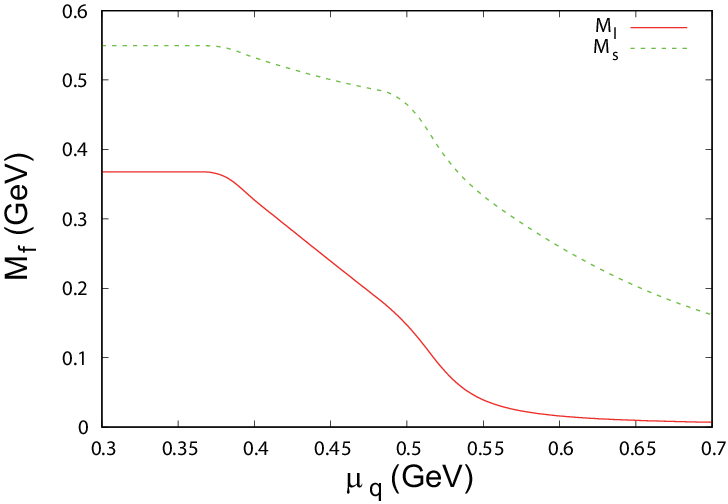}}
\caption{The $\mu_{\rm q}$-dependence of the effective quark mass $M_f$ with the crossover setup.  
The solid and dashed lines show $M_l~(l=u,d)$ and $M_s$, respectively. 
 }
 \label{Fig_Mf_010}
\end{figure}

Figure~\ref{Fig_Mf_010} shows the $\mu_{\rm q}$-dependence of the effective quark mass $M_f$. 
We see that $M_f$ starts to decrease when $\mu_{\rm q}$ exceeds the value of the effective light-quark mass at vacuum and gradually decreases as $\mu_{\rm q}$ increases. 
$M_f$ somewhat decreases early when $\mu_{\rm q}$ exceeds $0.5$ GeV. 
Comparing Fig.~\ref{Fig_volume_010} and Fig.~\ref{Fig_Mf_010}, we see that $M_f$ decreases early when $3v_{\rm q}$ is negligible. 

\begin{figure}[t]
\centering
\centerline{\includegraphics[width=0.40\textwidth]{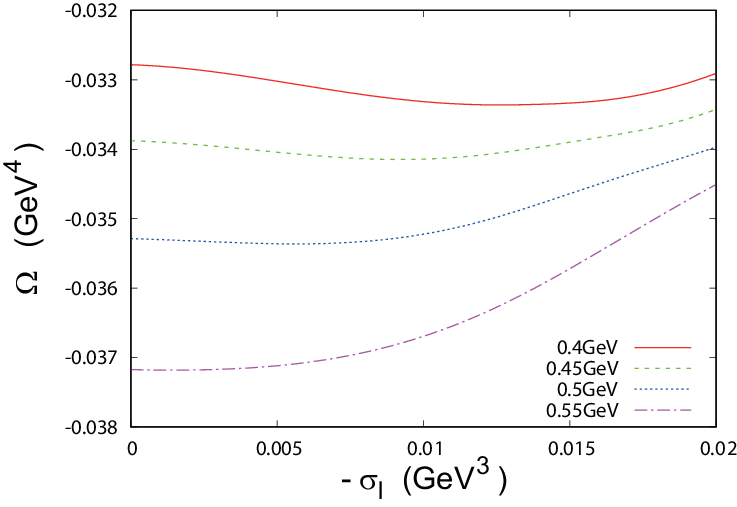}}
\caption{The $\sigma_l$-dependence of the thermodynamic potential $\Omega$ with the crossover setup.  
The solid, dashed, dotted and dot-dashed lines show the results at $\mu_{\rm q}=$0.4, 0.45, 0.5 and 0.55~GeV, respectively. 
For each case, the value of $\sigma_s$ is fixed to the value in the minimum of $\Omega (\sigma_f)$. 
}
 \label{Fig_Omega_c}
\end{figure}

With the crossover setup, the hybrid model approaches the NJL model gradually as $\mu_{\rm q}$ increases. 
Figure~\ref{Fig_Omega_c} shows the $\sigma_l~(l=u,d)$-dependence of the thermodynamic potential $\Omega$. 
We see that the value of $|\sigma_l|$ in the minimum of $\Omega$ decreases only slowly. 

\begin{figure}[h]
\centering
\centerline{\includegraphics[width=0.40\textwidth]{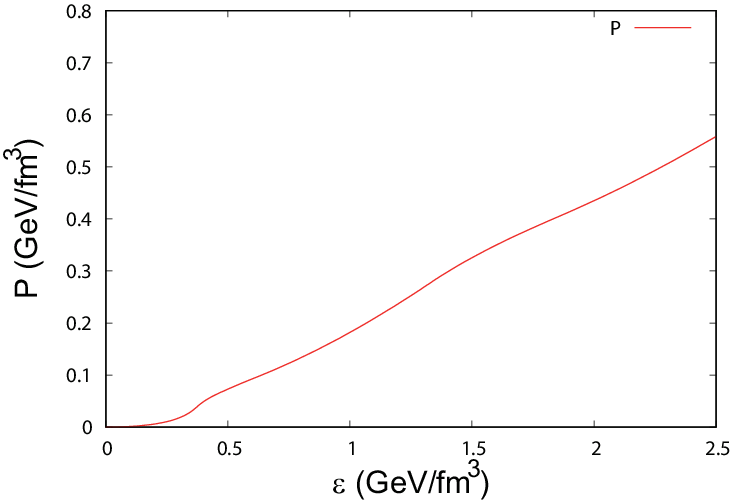}}
\caption{The relation between the energy density $\varepsilon$ and the pressure $P$ in our hybrid model with the crossover setup.}
 \label{Fig_e_p_010}
\end{figure}

Figure~\ref{Fig_e_p_010} shows the $\varepsilon$-$P$ relation. 
Since $\varepsilon$ and $P$ have the Dirac sea contributions which are not zero even at $\mu_{\rm q}=0$, we subtract $\varepsilon (\mu_{\rm q}=0)$ and $P(\mu_{\rm q}=0)$ from $\varepsilon$ and $P$, respectively.  
We see that $P$ increases monotonically as $\varepsilon$ increases. 
The intermediate phase mentioned above exists in the region $\varepsilon=0.2\sim 2.5$ GeV/fm$^3$. 

\begin{figure}[h]
\centering
\centerline{\includegraphics[width=0.40\textwidth]{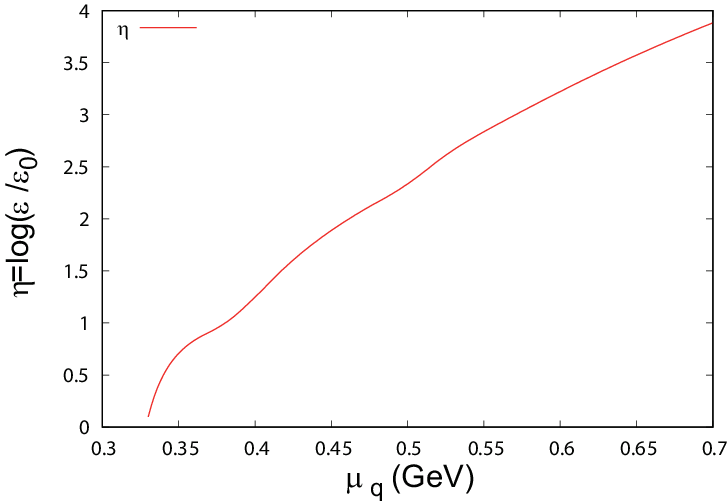}}
\caption{The $\mu_{\rm q}$-dependence of $\eta$ obtained by using our hybrid model with the crossover setup.  
 }
 \label{Fig_eta_010}
\end{figure}

\begin{figure}[h]
\centering
\centerline{\includegraphics[width=0.40\textwidth]{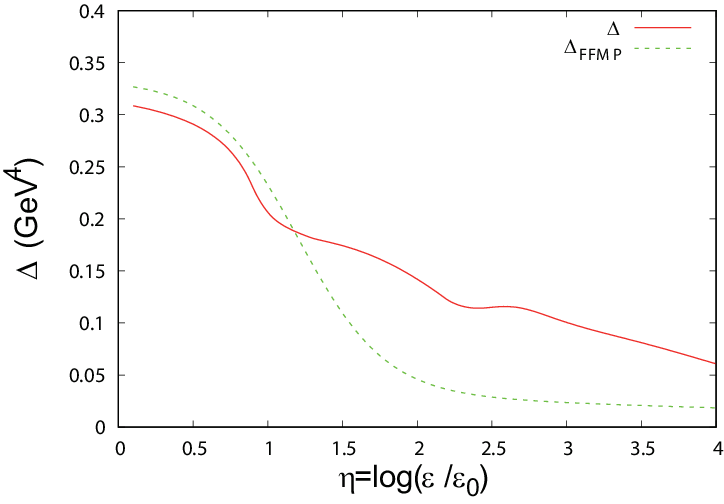}}
\caption{The $\eta$-dependence of $\Delta$ with the crossover setup.   
The solid and dashed lines show the result of our hybrid model and the result obtained by parameterization (7) in Ref.~\cite{Fujimoto:2022ohj}, respectively. 
 }
 \label{Fig_trace_010}
\end{figure}

According to Ref.~\cite{Fujimoto:2022ohj}, we introduce the following quantities, 
\begin{eqnarray}
\eta =\log{\left({\varepsilon\over{\varepsilon_0}}\right)}, 
\label{Eq_eta}
\end{eqnarray}
where $\varepsilon_0=0.15~{\rm GeV/fm^3}$, and 
\begin{eqnarray}
\Delta  ={1\over{3}}-{P\over{\varepsilon}}.  
\label{Eq_Delta}
\end{eqnarray}
When $\Delta$ vanishes, the trace anomaly $\varepsilon -3P$ becomes zero and the conformality is expected to be hold.  
Figure~\ref{Fig_eta_010} shows the $\mu_{\rm q}$-dependence of $\eta$.  
$\eta$ increases monotonically as $\mu_{\rm q}$ increases. 
The intermediate region $\mu_{\rm q}=0.33 \sim 0.55$ GeV corresponds to the region $\eta = 0.1 \sim 2.8$. 
The solid line in Fig.~\ref{Fig_trace_010} shows the $\eta$-dependence of $\Delta$ obtained by our hybrid model.  
The dashed line shows the result obtained by parameterization in Ref.~\cite{Fujimoto:2022ohj}, 
\begin{eqnarray}
\Delta ={1\over{3}}-{1\over{3}}\cdot {1\over{e^{-\kappa(\eta -\eta_c)}+1}}\left(1-{A\over{B+\eta^2}}\right), 
\label{FFMP7}
\end{eqnarray}
where $\kappa =3.45$, $\eta_c=1.2$, $A=2$ and $B=20$. 
In this paper, we call this parameterization $\Delta_{\rm FFMP}$. 
Our result $\Delta$ is in good agreement with $\Delta_{\rm FFMP}$ when $\eta <1$, but somewhat deviates from $\Delta_{\rm FFMP}$ at large $\eta$. 

\begin{figure}[h]
\centering
\centerline{\includegraphics[width=0.40\textwidth]{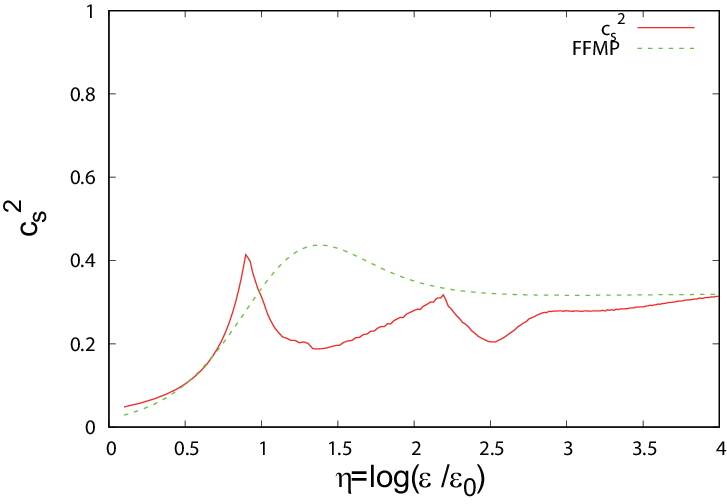}}
\caption{The $\eta$-dependence of $c_s^2$ with the crossover setup.  
The solid and dashed lines show the result of our hybrid model and the result obtained by parameterization (7) in Ref.~\cite{Fujimoto:2022ohj}, respectively. 
 }
 \label{Fig_cs2_010}
\end{figure}

\begin{figure}[h]
\centering
\centerline{\includegraphics[width=0.40\textwidth]{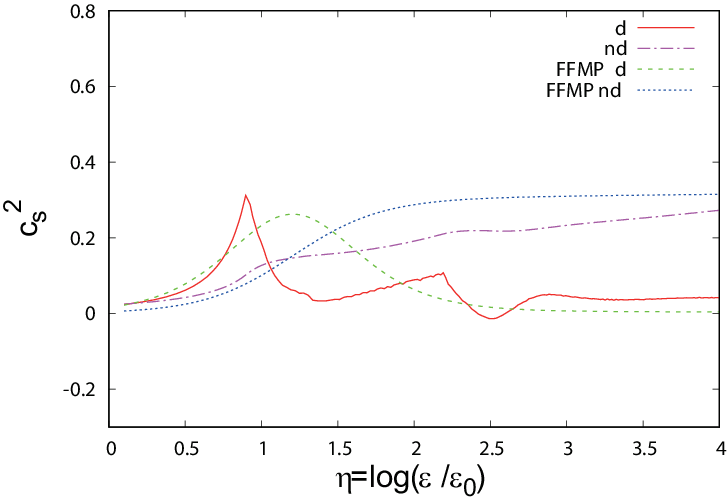}}
\caption{The $\eta$-dependence $c_{s,\rm d}^2$ and $c_{s,nd}^2$ with the crossover setup.   
The solid and dash-dotted lines show $c_{s,\rm d}^2$ and $c_{s,\rm nd}^2$ in our hybrid model, respectively. 
The dashed and dotted lines show $c_{s,\rm d}^2$ and $c_{s,\rm nd}^2$ obtained by parameterization (7) in Ref.~\cite{Fujimoto:2022ohj} , respectively. 
 }
 \label{Fig_cs2_d_010}
\end{figure}

Using $\Delta$ and $\eta$, the square of the speed of sound $c_s^2$ is rewritten as~\cite{Fujimoto:2022ohj} 
\begin{eqnarray}
c_s^2={dP\over{d\varepsilon}}=c_{s,\rm ,d}^2+c_{s,\rm nd}^2, 
\label{cs2_r}
\end{eqnarray}
where the derivative and non-derivative parts are given by 
\begin{eqnarray}
c_{s,d}^2=-{d\Delta\over{d\eta}},~~~~~~ c_{s,nd}^2={1\over{3}}-\Delta={P\over{\varepsilon}}.
\label{cs2_d_nd}
\end{eqnarray}
Figure~\ref{Fig_cs2_010} shows the $\eta$-dependence of $c_s^2$. 
$c_s^2$ has a double peak structure and, in the region $\eta=0.7\sim 3$, somewhat deviates from the parameterization result. 
The left peak is higher than the right one. 
It seems that the left peak is induced by the repulsion forces (i.e., excluded volume effects) among baryons, and the right one is related to the hadron-quark transition and the enhancement of the chiral symmetry restoration.  
It seems that this structure of $c_s^2$ corresponds to the right sketch in Fig.~2 in Ref.~\cite{Kojo:2020krb}. 
Figure~\ref{Fig_cs2_d_010} shows the $\eta$-dependence of $c_{s,\rm d}^2$ and $c_{s,\rm nd}^2$. 
It seems that the double peaks are originated in the derivative part. 
At large $\eta$, $c_{s,\rm d}^2$ and $c_{s,\rm nd}^2$ somewhat deviate from the parameterization results, but the deviations cancel each other, and the total $c_s^2$ is in good agreement with the parameterization. 

\subsection{First-order like transition}

\begin{figure}[h]
\centering
\centerline{\includegraphics[width=0.40\textwidth]{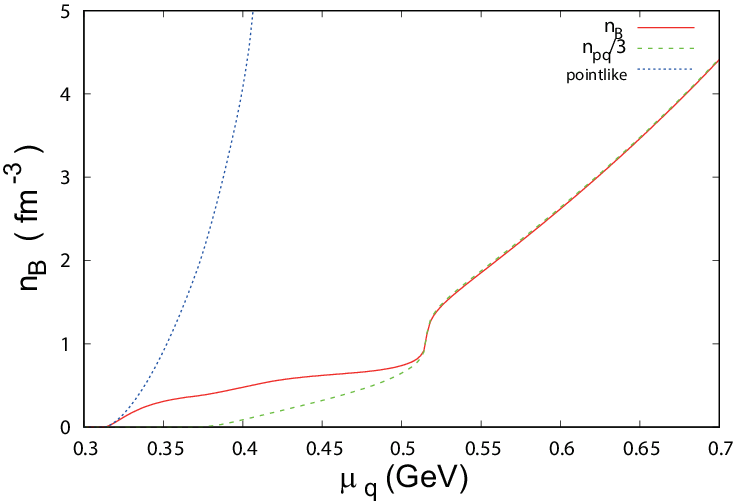}}
\caption{The $\mu_{\rm q}$-dependence of the baryon number density $n_{\rm B}$ 
with the first-order like setup.  
The solid, dashed, and dotted lines show $n_{\rm B}$, ${n_{\rm q}\over{3}}$ and $\tilde{n}_{\rm B}$ (pointlike), respectively. 
 }
 \label{Fig_nB_080}
\end{figure}

Next, we show the results with $a=0.8$; in the following, we call it the first-order like setup. 
Figure~\ref{Fig_nB_080} shows the $\mu_{\rm q}$-dependence of the baryon number density $n_{\rm B}$. 
We can see that $n_{\rm B}$ coincides with $\tilde{n}_{\rm B}$ at low density. 
As $\mu_{\rm q}$ increases, $n_{\rm B}$ approaches ${n_{\rm q}\over{3}}$. 
Finally, $n_{\rm B}$ coincides with ${n_{\rm q}\over{3}}$ at $\mu_{\rm q} =0.52$ GeV, and increases rapidly. 
In this case, it seems that the first-order-like transition occurs at $\mu_{\rm q} =0.52$ GeV. 
It is clear that the system is in a pure quark phase when $\mu_{\rm q} >0.52$ GeV. 
However, it seems that the system is in the intermediate phase rather than in the hadron phase in the region $\mu_{\rm q}=0.33\sim 0.52$ GeV. 

\begin{figure}[h]
\centering
\centerline{\includegraphics[width=0.40\textwidth]{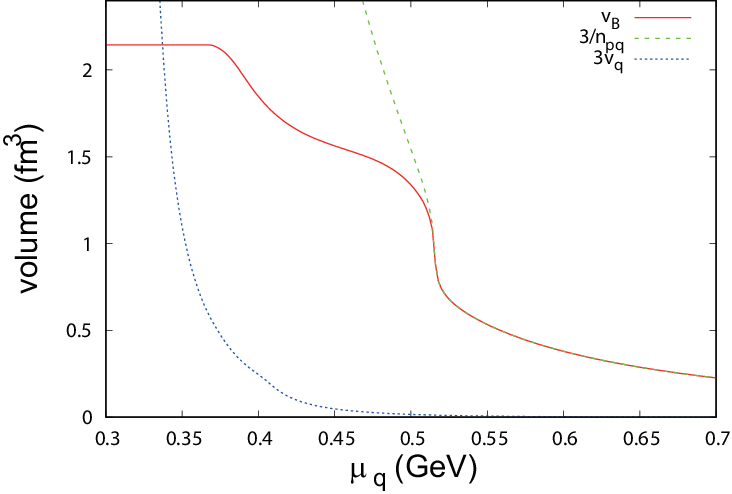}}
\caption{The $\mu_{\rm q}$-dependence of the baryon volume $v_{\rm B}$ with the first-order like setup.  
The solid, dashed, and dotted lines show $v_{\rm B}$, ${3\over{n_{\rm q}}}$ and $3v_{\rm q}$, respectively. 
 }
 \label{Fig_volume_080}
\end{figure}

Figure~\ref{Fig_volume_080} shows the $\mu_{\rm q}$-dependence of the baryon volume $v_{\rm B}$. 
We see that $v_{\rm B}$ coincides with $3/n_{\rm q}$ at $\mu_{\rm q}=0.52$ GeV and decreases rapidly.
When $\mu_{\rm q}>0.35$ GeV, $3v_{\rm q}$ is smaller than $v_{\rm B}$. 
Hence, it is natural to regard that the system is composed of quarks rather than baryons even when $\mu_{\rm q} =0.35\sim 0.52$ GeV, although it is not in the pure quark phase. 

\begin{figure}[h]
\centering
\centerline{\includegraphics[width=0.40\textwidth]{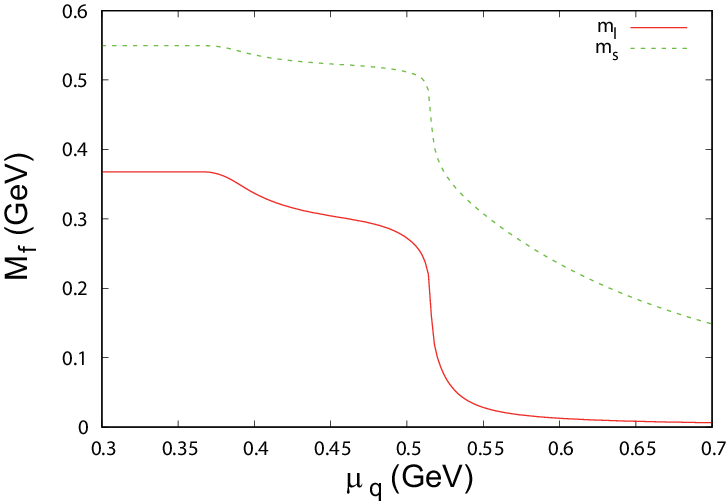}}
\caption{The $\mu_{\rm q}$-dependence of the effective quark mass $M_f$ with the first-order like setup.  
The solid and dashed lines show $M_l~(l=u,d)$ and $M_s$, respectively. 
 }
 \label{Fig_Mf_080}
\end{figure}

Figure~\ref{Fig_Mf_080} shows the $\mu_{\rm q}$-dependence of the effective quark mass $M_f$. 
We see that $M_f$ starts to decrease when $\mu_{\rm q}$ exceeds the value of the effective light quark mass at vacuum, gradually decreases as $\mu_{\rm q}$ increases, and then has an abrupt decrease at $\mu_{\rm q} =0.52$ GeV.  
The abrupt restoration of the chiral symmetry occurs at $\mu_{\rm q} =0.52$ GeV.  
It seems that this abrupt restoration of chiral symmetry is related to the abrupt increase of $n_{\rm B}$. 
It is known that in an abrupt transition, different physical quantities are correlated to each other near the transition point~\cite{Barducci:1992jm,Kashiwa:2009zz,Sakai:2010rp}.    
Hence, in this hybrid model, the abrupt changes in $n_{\rm B}$ and $M_{f}$ are correlated with each other. 

\begin{figure}[t]
\centering
\centerline{\includegraphics[width=0.40\textwidth]{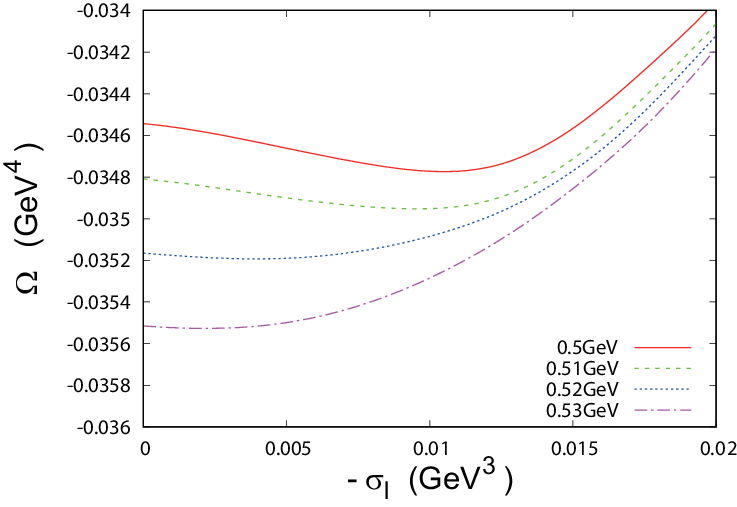}}
\caption{The $\sigma_l$-dependence of the thermodynamic potential $\Omega$ with the first-order like setup.  
The solid, dashed, dotted and dot-dashed lines show the results at $\mu_{\rm q}=$0.5, 0.51, 0.52 and 0.53~GeV, respectively. 
For each case, the value of $\sigma_s$ is fixed to the value in the minimum of $\Omega (\sigma_f)$. 
}
 \label{Fig_Omega_f}
\end{figure}

With the first-order like setup, the hybrid model approaches the NJL model abruptly as $\mu_{\rm q}$ increases. 
Figure~\ref{Fig_Omega_f} shows the $\sigma_l$-dependence of the thermodynamic potential $\Omega$. 
We see that the value of $|\sigma_l|$ in the minimum of $\Omega$ decreases rapidly. It should be noted that the bottom of $\Omega$ is almost flat at $\mu_{\rm q}=0.52$~GeV. 

\begin{figure}[t]
\centering
\centerline{\includegraphics[width=0.40\textwidth]{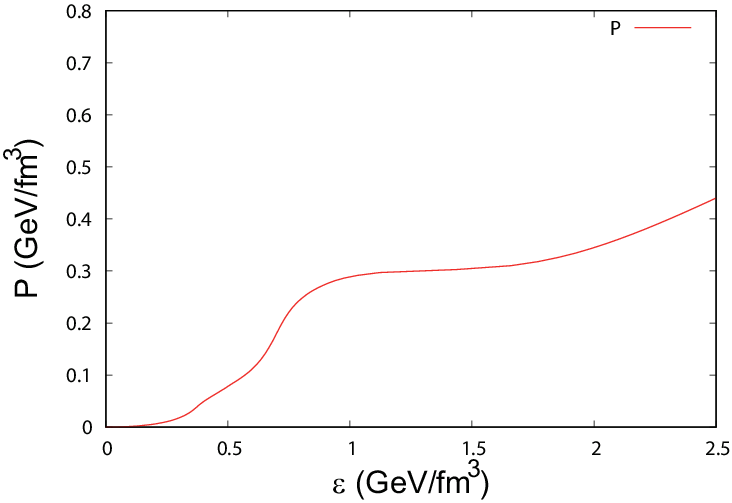}}
\caption{The relation between the energy density $\varepsilon$ and the pressure $P$ in our hybrid model with the first-order like setup.   
 }
 \label{Fig_e_p_080}
\end{figure}

Figure~\ref{Fig_e_p_080} shows the $\varepsilon$-$P$ relation. 
We see that $P$ increases monotonically as $\varepsilon$ increases when $\varepsilon >1~{\rm GeV/fm}^{3}$ or $\varepsilon <1.5~{\rm GeV/fm}^{3}$.  
There is a plateau of $P$ in the region $\varepsilon =1\sim 1.5~{\rm GeV/fm}^{3}$. 
This plateau is induced by the first-order like transition. 
The intermediate phase exists in the region $\varepsilon=0.2\sim 1.8$ GeV/fm$^3$. 

\begin{figure}[t]
\centering
\centerline{\includegraphics[width=0.40\textwidth]{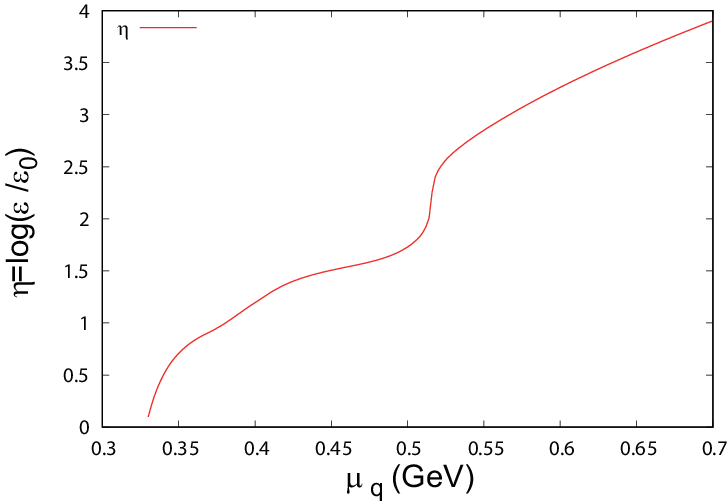}}
\caption{The $\mu_{\rm q}$-dependence of $\eta$ in our hybrid model with the first-order like setup.   
}
 \label{Fig_eta_080}
\end{figure}

Figure~\ref{Fig_eta_080} shows the $\mu_{\rm q}$-dependence of $\eta$.  
There is the tendency that $\eta$ increases monotonically as $\mu_{\rm q}$ increases and has an abrupt increase at $\mu_{\rm q} =0.52$ GeV. 
The intermediate region $\mu_{\rm q}=0.33 \sim 0.52$ GeV corresponds to the region $\eta = 0.1 \sim 2$.

\begin{figure}[t]
\centering
\centerline{~\includegraphics[width=0.40\textwidth]{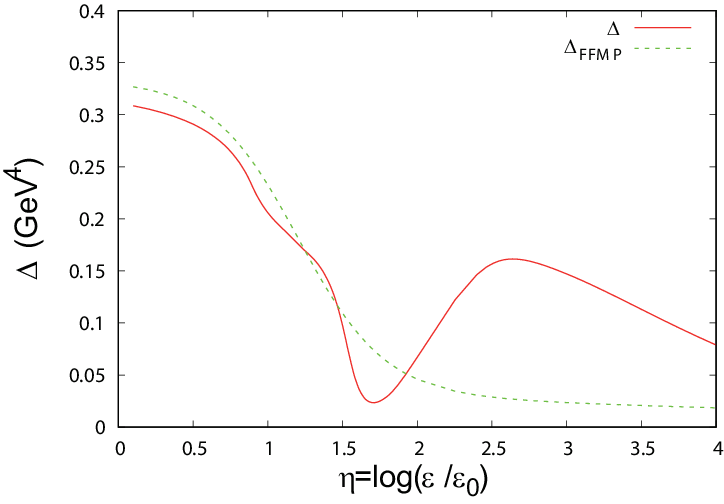}}
\caption{The $\eta$-dependence of $\Delta$ with the first-order like setup. 
The solid and dashed lines show the result of our hybrid model and the result obtained by parameterization (7) in Ref.~\cite{Fujimoto:2022ohj}, respectively. 
 }
 \label{Fig_trace_080}
\end{figure}

The solid line in Fig.~\ref{Fig_trace_080} shows the $\eta$-dependence of $\Delta$ in our hybrid model.  
Our result of $\Delta$ is in good agreement with $\Delta_{\rm FFMP}$, where $\Delta_\mathrm{FFMP}$ means the result obtained by parameterization (7) in Ref.~\cite{Fujimoto:2022ohj}, when $\eta <1.5$, but somewhat deviate from $\Delta_{\rm FFMP}$ at large $\eta$. 

\begin{figure}[t]
\centering
\centerline{\includegraphics[width=0.40\textwidth]{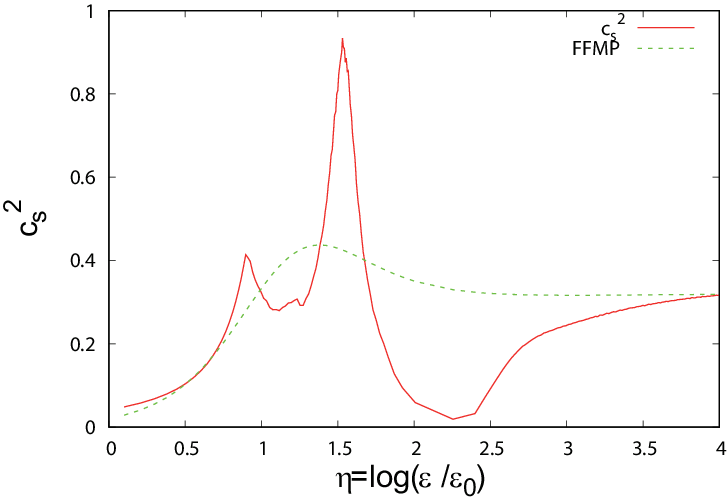}}
\caption{The $\eta$-dependence of $c_s^2$ with the first-order like setup.  
The solid and dashed lines show the result of our hybrid model and the result obtained by parameterization (7) in Ref.~\cite{Fujimoto:2022ohj}, respectively. 
 }
 \label{Fig_cs2_080}
\end{figure}

Figure~\ref{Fig_cs2_080} shows the $\eta$-dependence of $c_s^2$. 
We can see that $c_s^2$ has a double peak structure and, in the region $\eta=0.7\sim 3$, somewhat deviates from the parameterization result. 
In contrast to Fig.~\ref{Fig_cs2_010}, the right peak is higher than the left one.    
It seems that the left peak is induced by the repulsion forces among baryons and that the right peak is related to the hadron-quark transition and the chiral symmetry restoration.  
\begin{figure}[h]
\centering
\centerline{\includegraphics[width=0.40\textwidth]{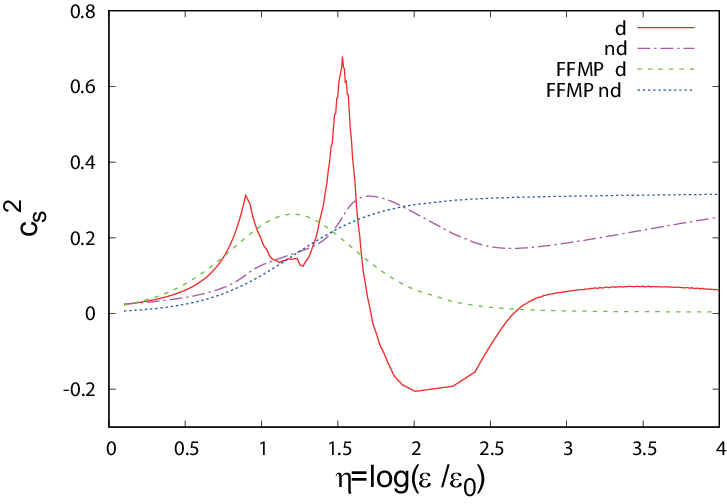}}
\caption{The $\eta$-dependence $c_{s,\rm d}^2$ and $c_{s,nd}^2$ with the first-order like setup.   
The solid and dash-dotted lines show $c_{s,\rm d}^2$ and $c_{s,\rm nd}^2$ in our hybrid model, respectively. 
The dashed and dotted lines show $c_{s,\rm d}^2$ and $c_{s,\rm nd}^2$ obtained by parameterization (7) in Ref.~\cite{Fujimoto:2022ohj}, respectively. 
 }
 \label{Fig_cs2_d_080}
\end{figure}
Figure~\ref{Fig_cs2_d_080} shows the $\eta$-dependence of $c_{s,\rm d}^2$ and $c_{s,\rm nd}^2$. 
It seems that the double peaks are originated in the derivative part. 
As in the case of Fig.~\ref{Fig_cs2_d_010}, $c_{s,\rm d}^2$ and $c_{s,\rm nd}^2$ somewhat deviate from parameterization results at large $\eta$, but the deviations cancel each other and the total $c_s^2$ is in good agreement with the parameterization.  
It seems that this structure of $c_s^2$ corresponds to the left sketch in Fig.~2 in Ref.~\cite{Kojo:2020krb}. 

\begin{figure}[h]
\centering
\centerline{\includegraphics[width=0.40\textwidth]{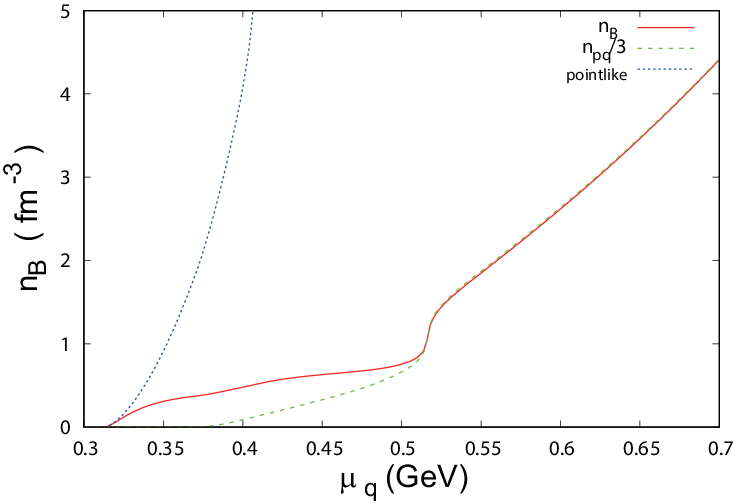}}
\caption{The $\mu_{\rm q}$-dependence of the baryon number density $n_{\rm B}$ 
with the baryon volume (\ref{volume_model2}).  
The solid, dashed, and dotted lines show $n_{\rm B}$, ${n_{\rm q}\over{3}}$ and $\tilde{n}_{\rm B}$ (pointlike), respectively. 
 }
 \label{Fig_nB_model2}
\end{figure}

\begin{figure}[t]
\centering
\centerline{\includegraphics[width=0.40\textwidth]{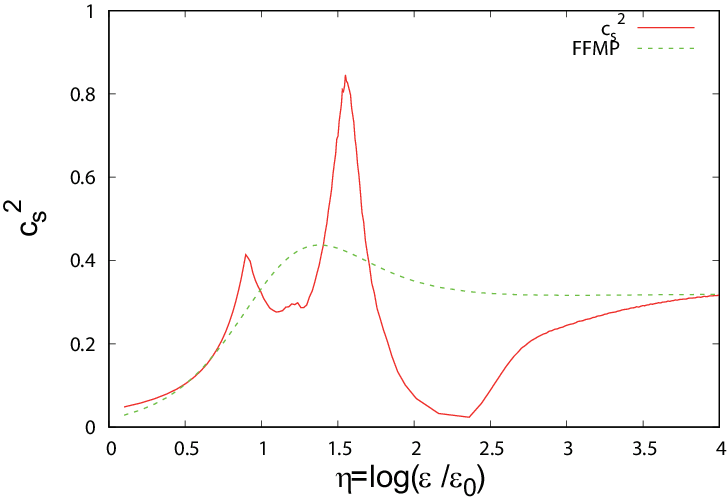}}
\caption{The $\eta$-dependence of $c_s^2$ with the baryon volume (\ref{volume_model2}).  
The solid and dashed lines show the result of our hybrid model and the result obtained by parameterization (7) in Ref.~\cite{Fujimoto:2022ohj}, respectively. 
 }
 \label{Fig_cs2_model2}
\end{figure}

It should be noted that the baryon number density has one to one correspondence to the other thermodynamic quantities since we only use the thermodynamic relations to calculate these quantities. 
Hence, the $\mu$-dependence of the baryon number density is strongly correlated with the $\mu$-dependence of the other thermodynamic quantities. 
Figure~\ref{Fig_nB_model2} shows the $\mu_{\rm q}$-dependence of the baryon number density when the baryon volume 
\begin{align}
v_{\rm B} &= {3\over{n_{\rm q}^\prime }},
\label{volume_model2}
\end{align}
with
\begin{align}
n_{\rm q}^\prime
&= n_{\rm q}
 + \frac{3}{{v_{\rm B0}}}
   \exp{ \Bigl[ -3\Bigl( \frac{v_{\rm B0} (n_{\rm q}+n_{\rm q}^3)}{6} \Bigr)^2 \Bigr]},
\label{pq_prime_model2}
\end{align}
is used instead of Eqs.~(\ref{volume_3}).  
Figure~\ref{Fig_nB_model2} resembles Fig.~\ref{Fig_nB_080}. 
Figure~\ref{Fig_cs2_model2} shows the $\eta$-dependence of the speed of sound. 
We see that Fig.~\ref{Fig_cs2_model2} also resembles Fig.~\ref{Fig_cs2_080}. 
In this meaning, the qualitative properties of the thermodynamic quantities do not depend on the detailed description of $v_{\rm B}$. 

\section{Summary}
\label{summary}

In summary, in this paper, the simple phenomenological hybrid model with the excluded volume effect (EVE) of baryons and the chiral dynamics is investigated.
In the model, we combine the Nambu--Jona-Lasinio model and the hadron resonance gas model.
The model interpolates between nuclear matter at saturation density and quark matter at high density. 
The equation of state (EoS) approaches that of pure quark matter as the density increases. 
There is the intermediate phase, which is difficult to be identified as a pure hadron phase or a pure quark phase. 
The speed of sound has a double-peak structure. 
One peak is related to the EVE of baryons, and the other is related to the hadron-quark transition and the chiral symmetry restoration.  
If the baryon volume $v_{\rm B}$ approaches ${3\over{n_{\rm q}}}$ gradually,  the hadron-quark transition is a typical crossover transition, where $n_{\rm q}$ is the quark number density of the pure quark phase.  
If the baryon volume $v_{\rm B}$ approaches ${3\over{n_{\rm q}}}$ abruptly,  the first-order like transition can occur in cooperation with the rapid chiral symmetry restoration, and the speed of sound can be very large. 

It seems that the density dependence of the baryon volume is very important for EoS at zero temperature. 
It is desirable to determine the dependence in the framework of the lattice QCD simulation. 
Simulations using the imaginary baryon chemical potential~\cite{deForcrand:2002hgr,DElia:2002tig,Nagata:2011yf,Takahashi:2014rta} may be useful for this purpose; see Refs.~\cite{Roberge:1986mm,Kashiwa:2019ihm} as an example. 

It is interesting to investigate neutron star properties using our hybrid model.
However, to extend the model to asymmetric matter, we should know not only the density dependence but also the isospin dependence of the baryon volume. 
The lattice QCD simulation with imaginary baryon and isospin chemical potentials~\cite{DElia:2009pdy} may be useful to determine the dependence. 

Recently, repulsion in nuclear matter is discussed~\cite{McLerran:2018hbz,Jeong:2019lhv, Duarte:2020kvi} in the context of quarkyonic matter~\cite{McLerran:2007qj}.  
In quarkyonic matter, the onset of the quark Fermi sea suppresses baryonic matter~\cite{Kojo:2021ugu}.  
The quarkyonic effective field theory~\cite{Duarte:2021tsx,Duarte:2023cki} is developed and the relation between the quarkyonic phase and EVE is discussed. 
Although the existence of quarkyonic matter is not explicitly assumed in our simple phenomenological hybrid model, there is some kind of quark-hadron duality and an intermediate phase which cannot be identified as pure hadron matter or pure quark matter appears.   
The study of the relation between quarkyonic matter and baryon volume may be important.  

On the thermal QCD transition, the possibility of the existence of the partial deconfinement phase has been discussed recently; for a recent review, see, e.g., Ref.~\cite{hanada2023thermal}. 
The partial deconfinement can be regarded as the coexistence of two phases in the internal color space rather than in the coordinate space. 
The intermediate phase which appears in our model may also be understood as such a phase.

\begin{acknowledgments}
K.K. is supported in part by the Grants-in-Aid for Scientific Research from JSPS (No. JP22H05112).
\end{acknowledgments}

\bibliography{ref.bib}

\end{document}